\documentclass{article}
\usepackage{longtable}
\usepackage{comment}
\usepackage{gensymb}
\usepackage[utf8]{inputenc} 
\usepackage[T1]{fontenc}    
\usepackage{hyperref}       
\usepackage{url}            
\usepackage{booktabs}       
\usepackage{amsfonts}       
\usepackage{nicefrac}       
\usepackage{microtype}      
\usepackage{lipsum}
\usepackage{graphicx}
\usepackage{amsmath,amssymb,amsfonts}
\usepackage{amsthm}
\usepackage{multirow}
\usepackage{multicol}

\usepackage{arxiv}

\author{Yantong Wang $^{1}$ and Vasilis Friderikos $^{2}$ \\
Center for Telecommunications Research\\
Department of Engineering, King's College London\\
London, WC2R 2LS, U.K. \\
$^{1}$ \texttt{yantong.wang@kcl.ac.uk}\\
$^{2}$ \texttt{vasilis.friderikos@kcl.ac.uk}\\
}
\title{A Survey of Deep Learning for Data Caching in Edge Network}

\begin{document}
\maketitle

\begin{abstract}
The concept of edge caching provision in emerging 5G and beyond mobile networks is a promising method to deal both with the traffic congestion  problem in the core network as well as reducing latency to access popular content. In that respect end user demand for popular content can be satisfied by proactively caching it at the network edge, i.e, at close  proximity to the users. In addition to model based caching schemes learning-based edge caching optimizations has recently attracted significant attention and the aim hereafter is to capture these recent advances for both model based and data driven techniques in the area of  proactive caching. This paper summarizes the utilization of deep learning for data caching in edge network. We first outline the typical research topics in content caching and formulate a taxonomy based on network hierarchical structure. Then, a number of key types of deep learning algorithms are presented, ranging from supervised learning to unsupervised learning as well as reinforcement learning. Furthermore, a comparison of state-of-the-art literature is provided from the aspects of caching topics and deep learning methods. Finally, we discuss research challenges and future directions of applying deep learning for caching.
\end{abstract}

\keywords{deep learning \and content caching \and network optimization \and edge network}

\section{Introduction}
\label{sec:introduction}
Undoubtedly, future 5G and beyond mobile communication networks will have to address stringent requirements of delivering popular content at ultra high speeds and low latency due to the proliferation of advanced mobile devices and data rich applications. In that ecosystem, edge-caching has received significant research attention over the last decade as an efficient technique to reduce delivery latency and network congestion especially during peak-traffic times or during unexpected network  congestion episodes by bringing popular data closer to the end users. One of the main reasons of enabling edge caching in the network is to reduce the number of requests that traverse the access and core mobile network as well as reducing the load at the origin servers that would have to, otherwise, respond to all requests directly in absence of edge caching. In that case popular content and objects can be stored and served from edge locations, which are closer to the end users. This operation is also beneficial from the end user perspective since edge caching can dramatically reduce the overall latency to access the content and increase in the sense overall user experience. It is also important to note that the notion of popular content means that the requests of top 10\% of video content on the Internet account for almost 80\% of all traffic; which relates to multiple requests from different end users of the same content
\cite{Infocom14}.

Recently, deep learning (DL) has attracted significant attention from both academia and industry and has been applied to diverse domains like self-driving, medical diagnosis, playing complex games such as Go \cite{sze2017efficient}. DL has also made their way into communication areas \cite{sun2019application}. In this paper, we pay attention to the application of DL in caching policy. Though there are some earlier surveys related to machine learning applications, they either focus on general machine learning techniques for caching \cite{kulkarni2020model,shuja2020applying,anokye2020survey}, or concentrate on overall wireless applications \cite{chen2019artificial,luong2019applications,wang2020convergence}. The work \cite{sun2019application} provides a big picture of applying machine learning in wireless communications. In \cite{kulkarni2020model}, the authors consider the machine learning on both caching and routing strategy. A comprehensive survey on machine learning applications for caching content in edge networks is provided in \cite{shuja2020applying}. The researchers \cite{anokye2020survey} provide a survey about machine learning on mobile edge caching and communication resources. On the other hand, \cite{chen2019artificial} overviews how artificial neural networks can be employed for various wireless network problems. The authors in \cite{luong2019applications} detail a survey on deep reinforcement learning (DRL) for issues in communications and networking. \cite{wang2020convergence} presents a comprehensive on deep learning applications and edge computing paradigm. Our work can be distinguished from the aforementioned papers based on the fact that we focus on the deep learning techniques on content caching and both wired and wireless caching are taken into account. Our main contributions are listed as follows:
\begin{itemize}
	\item We classify the content caching problem into Layer 1 Caching and Layer 2 Caching. Each layer caching consists of four tightly coupled subproblems: where to cache, what to cache, cache dimensioning and content delivery. Related researches are provided accordingly. 
	\item We present the fundamentals of DL techniques which are widely used in content caching, such as convolutional neural network, recurrent neural network, actor-critic model based deep reinforcement learning, etc.
	\item We analyze a broad range of state-of-the-art literature which use DL to content caching. These papers are compared based on the DL structure, layer caching coupled subproblems and the objective of DL in each scenarios. Then we discuss research challenges and potential directions for the utilization of DL in caching. 
\end{itemize}

\begin{figure}[htb]
	\centering
	\includegraphics[trim=0mm 0mm 0mm 0mm, clip, width=\textwidth]{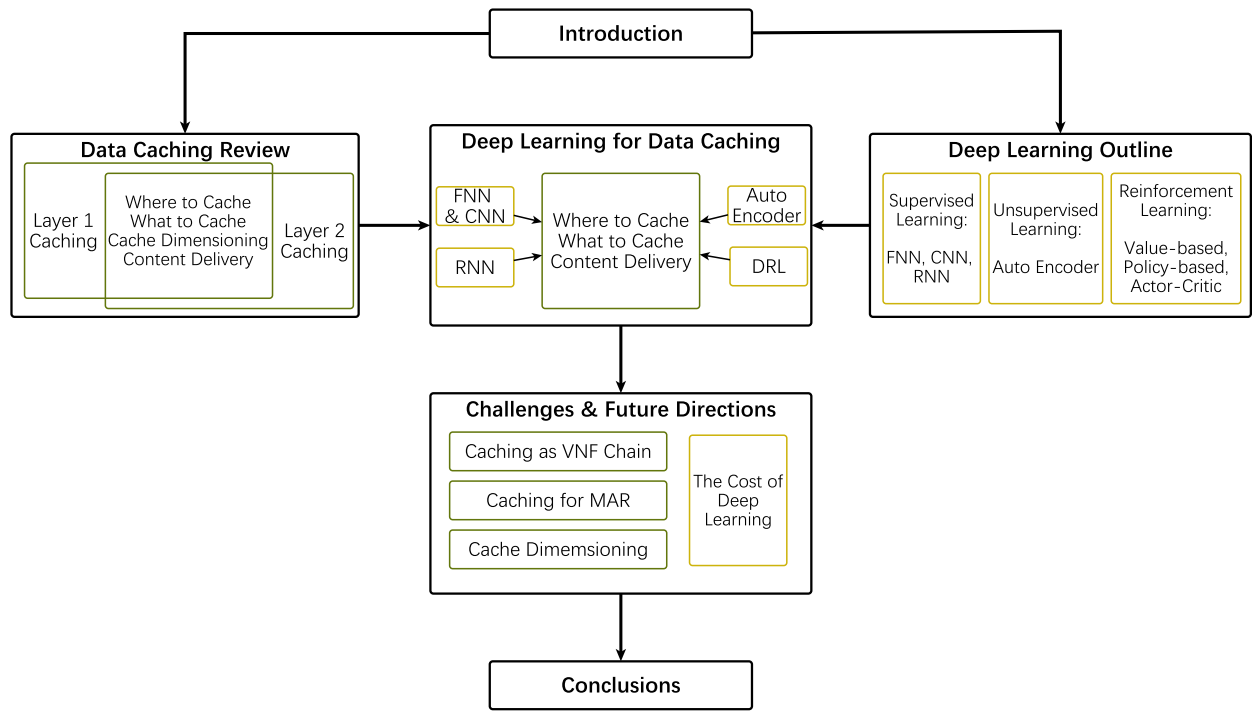}
	\caption{Survey Architecture}
	\label{fig:organization}
\end{figure}

The rest of this survey is organized as follows (as illustrated in Figure \ref{fig:organization} ). Section \ref{sec:caching} presents the categories of content caching problem. Section \ref{sec:dl} reviews typical deep neural network structures. In Section \ref{sec:dl_caching}, we list state-of-the-art DL-based caching strategies and their comparison. Section \ref{sec:challenges} debates challenges as well as potential research directions. In the end, Section \ref{sec:conclusions} concludes this paper. For better readability, the abbreviations in this paper is listed as Table \ref{tab:abbr} shows.

\begin{table}[htb]
	\centering
	\caption{\label{tab:abbr} List of Abbreviations.}
	\begin{tabular}{l|p{.38\textwidth}|l|p{.38\textwidth}}
		\hline
		\textbf{Abbr.} & \textbf{Description} & \textbf{Abbr.} & \textbf{Description} \\
		\hline
		3C & Computing, Caching and Communication & A3C & Asynchronous Advantage Actor-Critic \\
		\hline
		BBU & Baseband Unit & CCN & Content-Centric Network \\
		\hline
		CNN & Convolutional Neural Network & CoMP-JT & Coordinated Multi Point Joint Transmission \\
		\hline
		CR & Content Router & C-RAN & Cloud-Radio Access Network \\
		\hline
		CSI & Channel State Information & D2D & Device to Device \\
		\hline
		DDPG & Deep Deterministic Policy Gradient & DL & Deep Learning \\
		\hline
		DNN & Deep Neural Network & DQN & Deep Q Network \\
		\hline
		DRL & Deep Reinforcement Learning & DT & Digital Twin \\
		\hline
		ED & End Device & ES & Edge Server \\
		\hline
		ETSI & \multicolumn{3}{l}{European Telecommunication Standardization Institute} \\
		\hline
		ESN & Echo-State Network & FIFO & First In First Out\\
		\hline
		FNN & Feedforward Neural Network & FBS & Femto Base Station \\
		\hline
		ICN & Information-Centric Network & LFU & Least Frequently Used \\
		\hline
		LP & Linear Programming & LRU & Least Recently Used \\
		\hline
		LSTM & Long Short-Term Memory & MAR & Mobile Augmented Reality \\
		\hline
		MD & Mobile Device & MILP & Mixed Integer Linear Programming \\
		\hline
		MBS & Macro Base Station & NFV & Network Function Virtualization\\
		\hline
		PNF & Physical Network Function & PPO & Proximal Policy Optimization \\
		\hline
		QoE & Quality of Experience & RL & Reinforcement Learning \\
		\hline
		RNN & Recurrent Neural Network & RRH & Remote Radio Head \\
		\hline
		SAE & Sparse Auto Encoder & SDN & Software Defined Network \\
		\hline
		seq2seq & Sequence to Sequence & SNM & Shot Noise Model \\
		\hline
		TRPO & Trust Region Policy Optimization & TTL & Time to Live \\
		\hline
		VNF & Virtual Network Function & WSN & Wireless Sensor Network \\
		\hline
	\end{tabular}
\end{table}
\section{Data Caching Review}
\label{sec:caching}

\begin{figure}[htb]
    \centering
    \includegraphics[trim=1mm 1mm 5mm 1mm, clip, width=\textwidth]{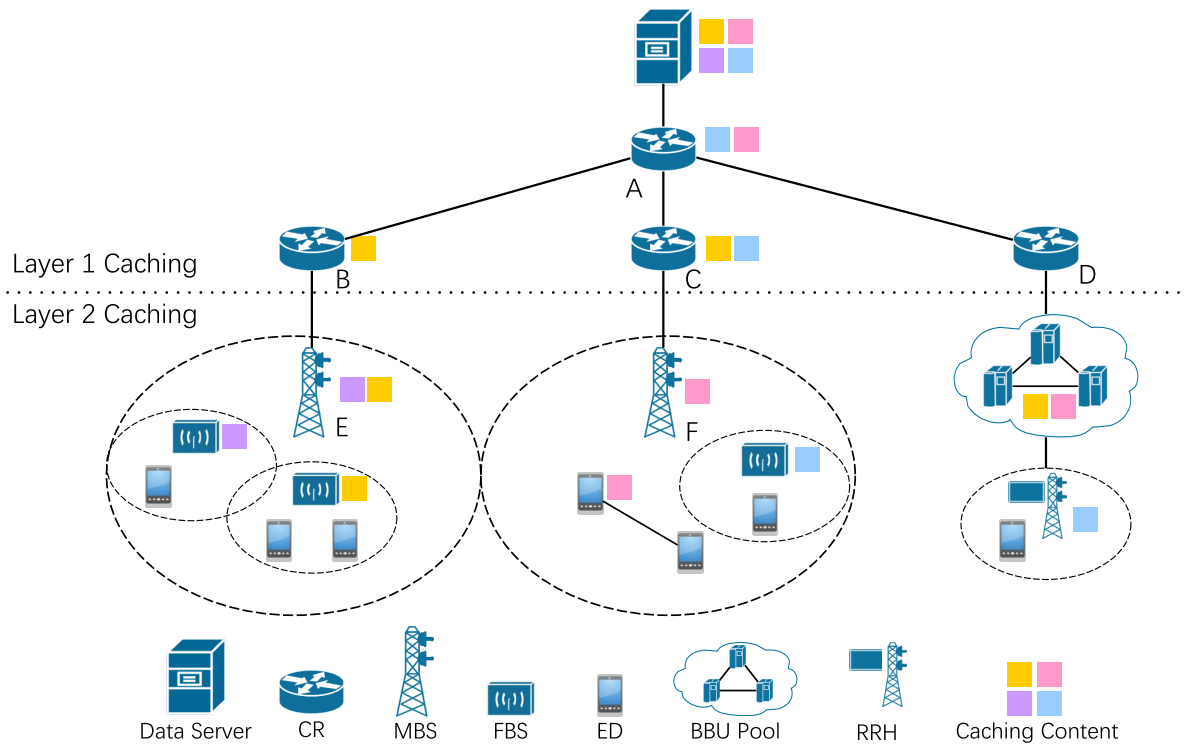}
    \caption{Data Caching in Edge Network}
    \label{fig:caching}
\end{figure}

The paradigm of data caching in edge networks is illustrated in Figure \ref{fig:caching}. Similarity to \cite{shi2016edge}, the scope of edge in this paper is along the path between end user and data server, which contains Content Router (CR), Macro Base Station (MSB), Femto Base Station (FSB) and End Device (ED). In the context of Cloud-Radio Access Network (C-RAN)\cite{peng2016recent}, both baseband unit (BBU) and remote radio head (RRH) are considered as potential caching candidates to hosting content, where the BBUs are clustered as a BBU pool centrally and RRHs are deployed near BS's antenna distributively. According to the hierarchical structure of edge network, the data caching is classified into two categories: Layer 1 Caching and Layer 2 Caching. In this section we illustrate the typical research topics in these two areas.

\subsection{Layer 1 Caching}
In Layer 1 Caching, the popular content is considered to be hosted in CRs. In the context of Information-Centric Network (ICN), CR plays dual roles both as a typical router (i.e. data flow forwarding) and content store (i.e. local area data caching facility). Generally, the CR is connected via wired networks. Layer 1 Caching consists of four tightly coupled problems: where to cache, what to cache, cache dimensioning and content delivery\cite{paschos2018role}.

Where to cache focuses on selecting the proper CRs to host the content. For instance, in Figure \ref{fig:caching}, contents replicas can be placed in lower hierarchical level CRs, such as router B and C, as the mean of reducing transmission cost but with extra cost pay for hosting contents; reversely, consolidating caching in CR A can be adopted to saving caching cost at the expense of more transmission cost and has the risk of expiring end users' delay requirement. Here the caching/hosting cost is the cost to deploy the content, which could be measured by space utilization, energy consumption or other metrics. The transmission cost represents the price for delivering the content from cached CR (or data server) to end user and is basically estimated via the number of hops. Where to cache problem usually has been modelled as a Mixed Integer Linear Programming (MILP):
\begin{subequations} 
\label{fml:MILP}
\begin{align}
\label{MILP:obj}
\mathop{\min}_{\substack{x}}\; & c^T x\\
\textrm{s.t.}\quad \label{MILP:con1}
& Ax\leq b \\
\label{MILP:con2}
& x\in\{0,1\} \\
\label{MILP:con3}
\text{or}\quad & x\geq 0
\end{align}
\end{subequations}
where $x$ is the decision variable. Normally it is a binary variable indicating the CR assignment. In special cases, with the aim of modelling or linearization, some non-binary auxiliary variables are introduced as constraint \eqref{MILP:con3} shows. If taking caching a part of a file not the complete into consideration, the decision variable $x$ is a continuous variable representing the segments host in the CR, then constraint \eqref{MILP:con2} becomes $x\in[0,1]$ and MILP model turns to linear programming (LP). There are many work allocating contents via MILP with different objectives and limitations. The authors in \cite{shan2019proactive} propose a model to minimize the user delay and load balancing level of CRs with the satisfaction of cache space. The work in \cite{wang2019proactive} considers a trade-off between caching and transmission cost with cache space, link bandwidth and user latency constraints. In \cite{fang2015energy}, an energy efficient optimization model is constructed consisting of caching energy and transport energy. \cite{sahoo2016survey} provides more details of mathematical model and related heuristic algorithms in caching deployment of wired networks. 

What to cache concentrates on selecting the proper contents in CRs for the purpose of maximizing the cache hit ratio. Via exploiting the statistical patterns of user requests, the popularity of requested information and user preference can be forecasted and play a very significant role in determining caching content. On the one hand, from the view of aggregated request contents, researchers propose many different models and algorithms for popularity estimation. One widely used model in web caching is the Zipf model based the assumption that the content popularity is static and each users' request is independent\cite{kabir2020role}. However, this method fails to reflect the temporal and spatial correlations of the content, where the temporal correlation reflects the popularity varies over time and the spatial correlation represents the content preference is different on the geographical area and social cultural media. A temporal model named the shot noise model (SNM) is built in \cite{traverso2013temporal} which enables users to estimate the content popularity dynamically. Inspired by SNM, the work in \cite{dabirmoghaddam2014understanding} considers both spatial and temporal characteristics during caching decision. On the other hand, from the view of a specific end user during a certain period, caching his/her preference content (may not be the popular in network) can also help to reduce the traffic flow. Many approaches in recommendation systems can be applied in this case\cite{shi2014collaborative}. Another aspect of what to cache problem is the designing of cache eviction strategies when storage space faces the risk of overflow. Depending on the life of caching contents, these policies can be divided into two categories roughly: one is like first in first out (FIFO), least frequently used (LFU), least recently used (LRU) and randomized replacement, the contents would not be removed until no more memory is available; the other one is called time to live (TTL) strategy, where the eviction happens once the related timer expires. \cite{jung2003modeling} presents analytic model for hit ratio in TTL-based cache requested by independent and identically distributed flows. It worth noting that in \cite{jung2003modeling}, the TTL-based cache policy is used for the consistency of dynamic contents instead of contents replacement. In \cite{fofack2014performance}, the authors introduce a TTL model for cache eviction and the timer is reset once related content cache hit happens.

Cache dimensioning highlights how much storage space to be allocated. Benefit from the softwarization and virtualization technologies, the cache size in each CR or edge cloud can be managed in a more flexible and dynamical way, which makes the cache dimensioning decisions become an important feature in data caching. Technically, the cache hit ratio rises with the increasing of cache memory, and consequently eases the traffic congestion in the core network. However, excessive space allocation would waste the resource like energy to support the caching function. Hence there is a trade-off between cache size cost and network congestion. Economically, taking such scenario into consideration: a small content provider wants to rent service from a CDN provider such as Akamai or Huawei Cloud, and there is also a balance between investment saving and network performance. In \cite{rossi2012sizing}, the proper cache size of individual CR in Content-Centric Network (CCN) is investigated via exploiting the network topology. In \cite{xu2014novel}, the authors consider the effect of network traffic distribution and user behaviours when designing cache size.

Content delivery considers how to transform the caching content to the requested user. The delivery traffic embraces single cache file downloading and video content steaming and the metrics for these two scenarios vary. Regarding file downloading, the content cannot be consumed until the delivery is completed. Therefore the downloading time of the entire file is viewed as a metric to reflect the quality of experience (QoE). For video steaming, especially for those large video splitted into several chunks, the delay limitation only works on the first chunk. In that case, delivering the first chunk in time and keep the smooth transmission of the rest chunks are the key aims \cite{paschos2019cache}. Apart from those measuring metrics, another problem in content delivery is the routing policy. In CCN \cite{jacobson2009networking}, one implementation of ICN architecture, employs a flooding-based name routing protocol to publish the request among cached CRs. On one hand, flooding strategy simplifies the designing complexity and reduce the maintaining cost particularly in an unstable scenario; on the other hand, it costly wastes bandwidth resources. In \cite{wang2018understanding}, the authors discuss the optimal radius in scoped flooding. The deliver route is often considered jointly with where to cache problem, in which the objective function \eqref{MILP:obj} includes both deployment and routing cost.

\subsection{Layer 2 Caching}
Contrast to Layer 1 caching in wired connection, Layer 2 caching considers implementing caching techniques in wireless network. Though both of them need solve where to cache, what to cache, cache dimensioning and content delivery problems, wireless caching is more challenging and some mature strategies in wired caching cannot be migrated directly to wireless case. Some reasons come from the listed aspects: the resources in wireless environment, such as caching storage and spectrum, are limited compared with CRs in Layer 1 Caching; the mobility of end users and dynamic network typologies are also required to be considered during the design of caching strategies; moreover, the wireless channels are uncertain since they can be effected by fading and interference.

In wireless caching, where to cache focus on finding the proper candidates among MBS, FBS, ED, even BBU pool and RRU in C-RAN to host the content. Caching at MBS and FBS can alleviate backhaul congestion since end users obtain the requested content from BS directly instead of from CR via backhaul links. Compared with FBS, MBS has wider coverage and typically, there is no overlap among different MBSs \cite{li2018survey}. As mentioned above, the caching space in BSs is limited and it is impractical to cache all popular content. With the aim of improving cache-hit ratio, a MILP-modelled collaborative caching strategy among MBSs is proposed in \cite{gharaibeh2015provably}. If the accessed MBS does not host the content, the request will be served by a neighbour MBS which cache the file rather than by the data server. For FBS caching, a distributed caching method is presented in \cite{golrezaei2013femtocaching} and the main idea is that the ED locating in the FBS coverage overlap is able to obtain contents from multiple hosters. Caching at ED can not only ease backhaul congestion but also improve the area spectral efficiency \cite{li2018survey}. When the end user requests a content, he/she would be severed by the local storage if the content is precached in his/her ED or by adjacent ED via D2D communication if the content is host accordingly. In \cite{afshang2016fundamentals}, the authors model the cache-enabled D2D network as a Poisson cluster process, where end users are grouped into several clusters and the collective performance is improved. Individually, caching the interested contents for other users affects personal benefit. In \cite{chen2016caching}, a Stackelberg game model is applied to formulate the conflict among end users and a related incentive mechanism is designed to encourage content sharing. For the case of cache-enabled C-RAN, caching at BBU can ease the traffic congestion in the backhaul while caching at RRH can reduce the fronthaul communication cost. On the other hand, caching all at BBU raises the signaling overhead of BBU pool while at RRH weakens the processing capability. Therefore, where to cache the content in C-RAN makes a substantial contribution to balancing the signal processing capability at the BBU pool and the backhaul/fronthaul costs \cite{li2018survey}. The work in \cite{ye2018tradeoff} investigates caching at RRHs with jointly considering cell outage probability and fronthaul utilization. Due to the end users' mobility, the prediction/awareness of user moving behaviour also influence the proper hoster selection. There are some researches exploiting user mobility in cache strategy designing like \cite{ren2020mobility} and \cite{song2019mobility}.

Similar with Layer 1, what to cache decision as well as eviction policy of layer 2 depends on the accurate prediction on content popularity or user preference in proactive caching method. The content popularity contains the feature of temporal and spatial correlations, which has already been described in Layer 1 Caching. In Layer 2 caching, the proper spatial granularity in popular contents estimation needs to take special attentions \cite{liu2016caching}. For example, the coverage of MBS and FBS are different, which makes the popularity in MBS and FBS are different as well. Because the former based on a large number of users' behaviors but the individual may prefer specific content categories. For small cells, the preference estimation requires more accurate information like historical data \cite{li2018survey}. In order to capture the temporal and spatial dynamics of user preference, many different deep learning based algorithms are proposed, which will be illustrated in Section \ref{sec:dl_caching}.

Cache dimensioning in Layer 2 Caching has more complicated factors need to be considered, not only including the network topology and content popularity as Layer 1 Caching, but also containing backhaul transmission status and wireless channel features. The proper cache size assignment is studied in the scenario of backhaul limited cellular network \cite{peng2016cache}. It also provides the closed-form boundary of minimum cache size in one cell case. In the case of dense wireless network, the work in \cite{liu2016much} quantifies the minimum required cache to achieve the linear capacity scaling of network throughput. The authors of \cite{song2017minimum} also consider the scenario of dense networks. They derive the closed-form of the optimal memory size which can reduce the consumption of backhaul capacity as well as guarantee wireless QoS.

According to the number of transmitters and receivers, we divide the content delivery in Layer 2 caching into three categories: one candidate serves one end user, such as unicast and D2D transmission; one candidate serves multiple users like multicast; and coordinated delivery including multiple transmitters serve one or more receivers like coordinated multi-point joint transmission (CoMP-JT). Once the requested content is cached locally, BS can serve the end user via unicast or the adjacent device shares the contents by implementing D2D transmission. Concurrent transmission has the risk of co-channel interference in dense deployed networks. In D2D network, link scheduling is introduced to select subsets of links to transmit simultaneously \cite{li2018survey}. With the aim  of improving the spectral efficiency, multicast is applied in content delivery when serving multiple requests simultaneously with the same content. Therefore there is a trade off between spectral efficiency and service delay. For the aim of serving more users in one transmission as well as higher spectral efficiency, the BS will wait to collect enough requirement for the same content which makes the first request a long waiting time. An optimal dynamic multicast scheduling is proposed in \cite{zhou2017optimal} to balance these two factors. Multicast can also serve multiple requests with different contents. In \cite{maddah2014fundamental}, the authors provides a coded caching scheme which requires the communication link is error free and each user caches a part of its own content and partial of other users. Then BS multicasts the coded data to all users. Each user can decode his own requested content by XOR operation between the received data and the precached other users' file. However, the coding complexity increases exponentially as the quantity of end users grows. The CoMP-JT can improve the spectral efficiency as well via sharing channel state information (CSI) and contents among BSs but it also needs high-capacity backhaul consumption for exchanging data. In C-RAN, the BBUs are centralized in the BBU pool, which makes communication among BSs very efficiency. \cite{ha2015coordinated} designs CoMP-JT in C-RAN for the purpose of minimizing power consumption with limitations of transmission energy, link capacity and requested QoS.
\section{Deep Learning Outline}
\label{sec:dl}

\begin{figure}[htb]
    \centering
    \includegraphics[trim=0mm 0mm 0mm 0mm, clip, width=\textwidth]{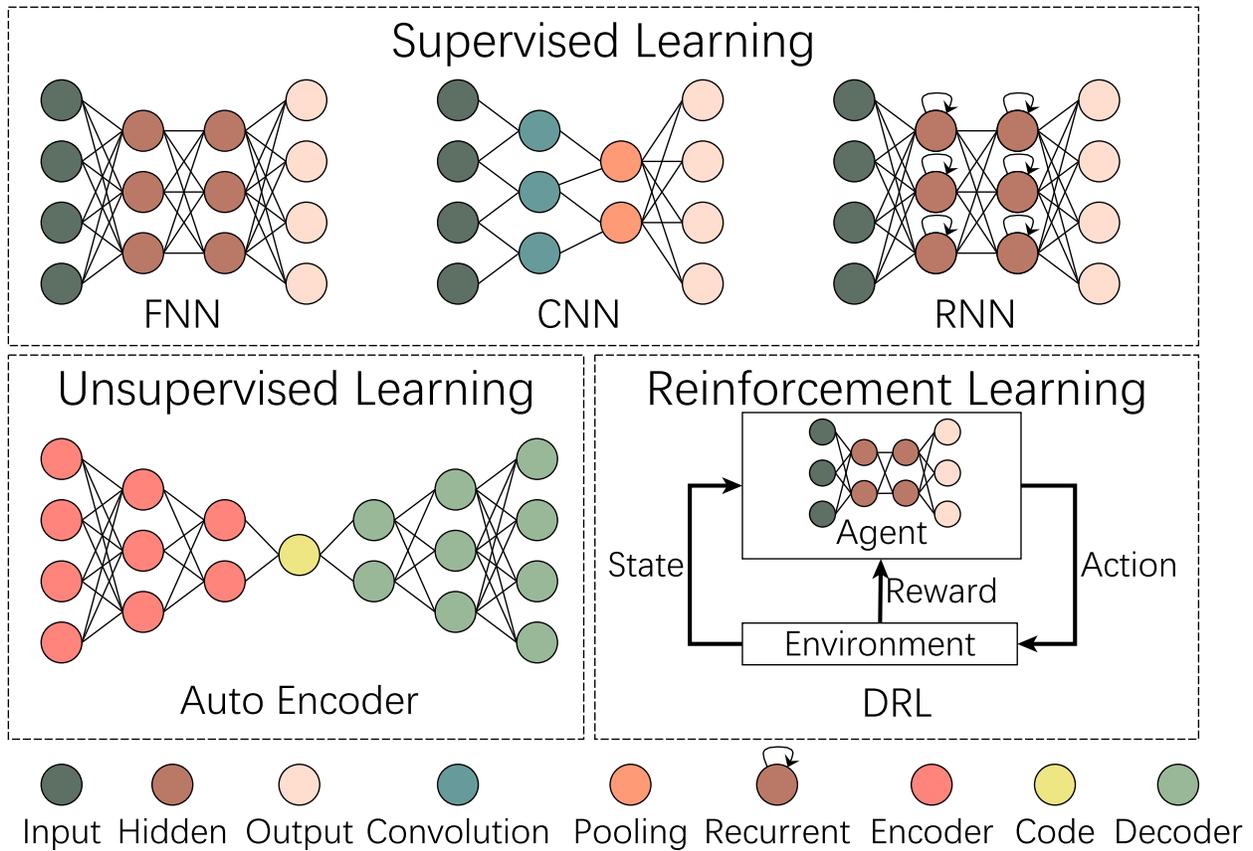}
    \caption{Typical DNN Structures}
    \label{fig:DNN}
\end{figure}

As Figure \ref{fig:DNN} shows, some typical deep neural network (DNN) methods are stated. These models are classified into three categories depending on the training methods: supervised learning, unsupervised learning and reinforcement learning. 

\subsection{Feedforward Neural Network (FNN)}
FNN is a kind of DNNs whose information propagation direction is forward and there is no cycle in neurons. In this paper, the term FNN is used to represent fully connected neural network, which indicates the connection between two adjacent layers is filled. According to the Universal Approximation Theorem, FNN has the ability to approximate any closed and bounded function with enough neurons in hidden layer \cite{goodfellow2016deep}. The hidden layer is applied to extract features of input vector, and then feed the output layer, which works as a classifier. Though FNN is very powerful, it gets into trouble when dealing with real-world task such as image recognition due to enormous weight parameters (because of fully connected) and lack of data augmentation.   

\subsection{Convolutional Neural Network (CNN)}
For the aim of overcoming the aforementioned drawback of FNN, CNN employs convolution and pooling operations, where the former applies sliding convolutional filters to the input vector and the later does down sampling, usually via maximum or mean pooling. Generally, CNN tends to contain deeper layers and smaller convolutional filters, and the structure becomes fully convolutional network \cite{long2015fully}, reducing the ratio of pooling layers as well as fully connected layers. Taxonomically, CNN belongs to FNN and has been broadly employed in image recognition, video analysis, natural language processing, etc. Including CNN, one of the limitations of FNN is that the output only depends on current input vectors. So it is hard to deal with sequential tasks.

\subsection{Recurrent Neural Network (RNN)}
\begin{figure}[htb]
    \centering
    \includegraphics[trim=0mm 0mm 35mm 0mm, clip, width=\textwidth]{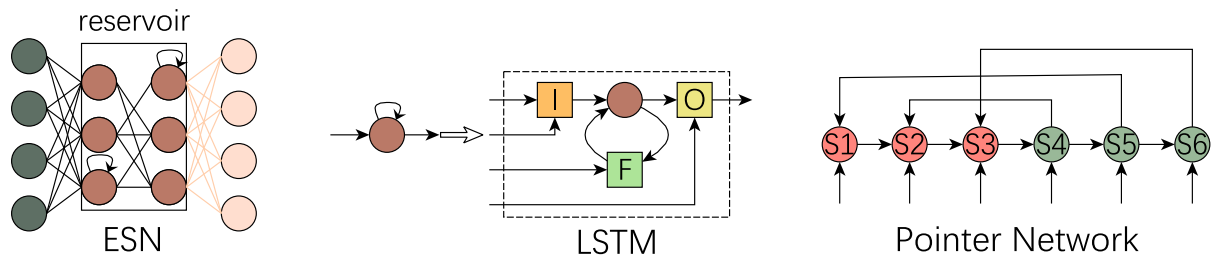}
    \caption{RNN Variants}
    \label{fig:RNN}
\end{figure}

In order to deal with sequential tasks and using historical information, RNN employs neurons with self feedback in hidden layers. Unlike the hidden neuron in FNN, the output of recurrent neuron depends on both current output of precious layer and last hidden state. Compared with FNN approximates any continues functions, RNN with Sigmoid activation function can simulate a universal Turing Machine and has the ability to solve all computational problems \cite{siegelmann1991turing}. It is worth noting that RNN has the risk to suffer from long-term dependencies problem \cite{goodfellow2016deep} including gradient exploding and vanishing. Additionally, RNN has more parameters waiting to be trained due to adding recurrent weights. In the following, we introduce some RNN variants as Figure \ref{fig:RNN} shows.

\subsubsection{Echo-State Network (ESN)}
As aforementioned, simple RNN contains more parameters in training step, where the recurrent weights and input weights are difficult to learn \cite{goodfellow2016deep}. The basic idea of ESN is fixing these two kinds of weights and only learn the output weights (as links highlighted in Figure \ref{fig:RNN}). The hidden layer is renamed as reservoir in ESN, where the neurons are sparsely connected and the weights are randomly assigned. The recurrent weights keep constant so the information of previous moments is stored in the reservoir with constant weight like voice echoing. 

\subsubsection{Long Short-Term Memory (LSTM)}
Recently, an efficient way to cope with long-term dependencies in practical is employing gated RNN, including LSTM \cite{goodfellow2016deep}. Then we compare with the recurrent neuron in simple RNN: Internally LSTM introduces three gates to control signal propagation, where input gate $I$ decides the partition of input signal to be stored, forget gate $F$ controls ratio of last moment memory to be kept until next period (the name "forget gate" may be a little misleading because it actually represents the ratio to be remembered) and output gate $O$ influences the proportion of current state to be delivered; Externally LSTM has four inputs embracing one input signal and three control signals for three gates. All these four signals are derived via the calculation of current network input and last moment delivered state.

\subsubsection{Pointer Network}
A typical application of RNN is converting one sequence to another sequence (seq2seq) such as machine translation. Conventionally, the output of seq2seq architecture is a probability distribution of output dictionary. However, it cannot deal with the problem that the size of output relies on the length of input due to fixed output dictionary. In \cite{vinyals2015pointer}, the authors modify the output to be the distribution of input sequence, which is analogous to pointers in C/C++. Pointer network has been widely used in text condensation.

\subsection{Auto Encoder}
Auto Encoder is a stack of two NNs named encoder and decoder respectively, where the former tries to learn the representative characteristics of input and generate a related code, and the later reads the code and reconstructs the original input. In order to avoid the auto encoder simply copying the input, some restrictions are considered like the dimension of code is smaller than input vector \cite{goodfellow2016deep}. The quality of auto encoder can be measured via reconstruction error, which estimates the similarity between input and output. In most cases, the auto encoder is used for the proper representation of input vector so the decoder part is removed after unsupervised training. The code can be employed as input for further deep learning models.

\subsection{Deep Reinforcement Learning (DRL)}
Reinforcement Learning (RL) is a Markov Decision Process represented by a quintuple $\{\mathcal{S,A,P,R},\gamma\}$, where $\mathcal{S}$ is the state space controlled by environment; $\mathcal{A}$ is the action space determined by agent; $\mathcal{P}$ is the state transition function measuring the probability of moving to a new state $s_{t+1}$ given previous state $s_t$ and action $a_t$; $R$ is reward function calculated by environment considering state and action; $\gamma$ is a discount factor for estimating total reward. During the interaction between agent and environment, agent observes current state $s_t$ from environment, and then takes action $a_t$ following its policy $\pi$. The environment moves to a new state $s_{t+1}$ stochastically based on $\mathcal{P}(s_t,a_t)$ and returns a reward $r_t$ to agent. The RL's aim is finding the policy $\pi$ to maximum accumulated reward $\sum_t\gamma^tr_t$. In the early stage, RL focuses on scenarios whose $\mathcal{S}$ and $\mathcal{A}$ are discrete and limited. So the agent can use a table to record these information. Recently, some tasks have enormous discrete states and actions such as playing go and even continuous value such as self-driving, which makes table recording impractical. In order to solve this, DRL combines RL and DL, where RL defines the problem and optimization object; DL models the policy and the reward expectation. Depending on the roles of DNN in DRL, we classify the DRL into 3 categories as Figure shows.

\subsubsection{DNN as Critic (Value-Based)}
In value-based method, DNN does not get involved with policy decision but estimates the policy performance. Two functions are introduced for the measurement: $V^\pi(s)$ represents the reward expectation of policy $\pi$ starting from state $s$; $Q^\pi(s,a)$ illustrates the reward expectation of policy $\pi$ starting from state $s$ and taking action $a$. In addition, $V^\pi(s)$ is the expected value of $Q^\pi(s,a)$. If we can estimate $Q^\pi(s,a)$, the policy $\pi$ can also be improved by choosing the action $a^*$ hold $Q^\pi(s,a^*)\geq V^\pi(s)$. So the DNN employed in agent is approximating function $Q^\pi(s,a)$, where the inputs are state $s$ and action $a$ and output is the estimated value $Q^\pi(s,a)$. There are some representative critic methods like Deep Q Networks (DQN) \cite{mnih2015human} and its variants Double DQN \cite{van2016deep}, Dueling DQN \cite{wang2016dueling}, etc.

\subsubsection{DNN as Actor (Policy-Based)}
In policy-based method, DNN gets involved in the action selection directly instead of via $Q^\pi(s,a)$. The policy can be viewed as an optimization problem, where the objective function is maximizing reward expectation and the search space is policy space. The input of DNN is current state and output is the probability distribution of potential actions. By employing gradient ascent, we can update the DNN to provide better action then maximize total reward. Some popular algorithms include Trust Region Policy Optimization (TRPO) \cite{schulman2015trust}, Proximal Policy Optimization (PPO) \cite{schulman2017proximal}.

\subsubsection{Actor-Critic Model}
Generally, compared with policy-based approach, the value-based method is less stable and suffer from poor convergence since the policy is derived based on $Q^\pi(s,a)$ approximation. But value-based method is more sample efficient, while policy-based method is easier to fall into local optimal solution because the search space is vast. The actor-critic model combines these two approaches, i.e. the agent contains two DNNs named actor and critic respectively. In each training iteration, the actor considers current state $s$ and policy $\pi$ for deciding action $a$. Then the environment changes to state $s'$ and returns reward $r$. The critic updates its own parameters based on the feedback from environment and output a mark for the actor's action. The actor updates the policy $\pi$ depending on critic's mark. Some typical algorithms are proposed recent years like Deep Deterministic Policy Gradient (DDPG) \cite{lillicrap2015continuous} and Asynchronous Advantage Actor-Critic (A3C) \cite{mnih2016asynchronous}.
\section{Deep Learning for Data Caching}
\label{sec:dl_caching}

We divide the studies regarding deep learning for data caching in edge networks into four categories depending on the DL tools employed: FNN and CNN; RNN; Auto Encoder; DRL. Recently many works utilize more than one DL techniques for jointly considered caching problems. For instance, at the beginning we applies a RNN to predict content popularity, and then a DRL to find suboptimal solutions of content placement for the purpose of reducing time complexity. In such case, we classify the related work into DRL since it represents the caching allocation policy. Unless mention the caching location (such as CRs, MBSs, FBSs, EDs and BBUs) otherwise, the approaches in this section can be utilized for both Layer 1 and Layer 2 caching. Table \ref{tab:summary} summarize some studies of DL for caching.

\begin{table}[htb]
\caption{Summary of Deep Learning for Data Caching}
\label{tab:summary}
\begin{tabular}{p{.07\textwidth}|p{.1\textwidth}|p{.28\textwidth}|p{.43\textwidth}}
    \hline
    \textbf{method} & \textbf{Study} & \textbf{Caching Problem} & \textbf{DL Objective}\\
    \hline
    \multirow{8}{.07\textwidth}{FNN and CNN} & \cite{lei2019learning} & content delivery & reduce feasible region of time slot allocation\\
    \cline{2-4}
    & \cite{lei2017deep} & where to cache, content delivery & determine MBSs for caching \& delivery duration \\
    \cline{2-4}
    & \cite{wang2019caching} & where to cache, content delivery & nominate proper CRs for caching \\
    \cline{2-4}
    & \cite{wang2020network} & where to cache, content delivery & reduce feasible region for caching \\
    \cline{2-4}
    & \cite{doan2018content} & what to cache & extract video features\\
    \cline{2-4}
    & \cite{qin2019neural} & what to cache & predict requested content \& frequency \\
    \cline{2-4}
    & \cite{tsai2018mobile} & what to cache & predict requested content \\
    \cline{2-4}
    & \cite{fedchenko2019feedforward} & what to cache & predict content popularity\\
    \hline
    \multirow{4}{.07\textwidth}{RNN} & \cite{chen2017echo,chen2017caching} & what to cache & predict requested content \& user mobility \\
    \cline{2-4}
    & \cite{ale2019online,fan2020pa,zhang2019toward,mou2019lstm,narayanan2018deepcache} & what to cache & predict content popularity \\
    \cline{2-4}
    & \cite{zhang2019cover} & content delivery & reduce traffic load, select optimal BS subset\\
    \hline
    \multirow{3}{.07\textwidth}{Auto Encoder} & \cite{lei2019deep,lei2018proactive,rathore2019deepcachnet,liu2017content,li2019reinforcement} & what to cache & predict content popularity \\
    \cline{2-4}
    & \cite{lin2020video} & what to cache & predict top popular contents \\
    \hline
    \multirow{17}{.07\textwidth}{DRL}  &\cite{zhong2018deep,zhong2020deep,gao2020reinforcement,maniotis2020viewport,he2019qoe,tang2019energy,wu2019dynamic,zhu2018caching,sadeghi2019deep,wang2019deepchunk} & what to cache & decide cache placement \\
    \cline{2-4}
    & \cite{rahman2020learning} & what to cache & decide cache replacement \& power allocation \\
    \cline{2-4}
    & \cite{thar2019deep} & what to cache & predict popularity \& searching best NN model \\
    \cline{2-4}
    & \cite{he2018trust} & where to cache & decide cache location \\
    \cline{2-4}
    & \cite{he2017deep} & content delivery & users grouping\\
    \cline{2-4}
    & \cite{he2017integrated} & where to cache, content delivery & decide BS connection, computation offloading \& caching location\\
    \cline{2-4}
    & \cite{qiao2019deep} & what to cache, content delivery & decide caching \& bandwidth allocation \\
    \cline{2-4}
    &\cite{wei2018joint} & what to cache, content delivery & decide caching, computing offloading \& radio resource allocation \\
    \cline{2-4}
    & \cite{zhang2019double} & what to cache, content delivery & decide multicast scheduling \& caching replacement\\
    \cline{2-4}
    & \cite{li2020joint} & where \& what to cache & predict popularity, decide caching \& task offloading \\
    \cline{2-4}
    & \cite{li2019deep} & where \& what to cache, content delivery & predict user mobility \& content popularity, determine D2D link\\
    \hline
\end{tabular}
\end{table}

\subsection{FNN and CNN}
In \cite{lei2019learning}, the content delivery problem in wireless network is formulated as two MILP optimization models with the aims of minimum delivery time slot and energy consumption respectively. Both models consider the data rate for content delivery. Considering the computational complexity of solving MILP, a CNN is introduced to reduce the feasible region of decision variables, where the input is channel coefficients matrix. The FNN in paper \cite{lei2017deep} plays a similar role as \cite{lei2019learning} to simplify the searching space of the content delivery optimization model. 

For resource allocation problem, the authors of \cite{lee2018deep} model it as linear sum assignment problems then utilize CNN and FNN to solve the model. The idea is extended in \cite{wang2019caching} and \cite{wang2020network}, where the authors consider where to cache problem among potential CRs and content delivery jointly, which is modeled as MILP with the aim of balancing caching and transmission cost by considering the user mobility, space utilization and bandwidth limitations. The cache allocation is viewed as multi-label classification problem and is decomposed into several independent sub-problems, where each one correlates with a CNN to predict assignment. The input of CNN is a grey-scale image which combines the information of user mobility, space and link utilization level. In \cite{wang2019caching}, a hill climbing local search algorithm is provided to improve the performance of CNN while in \cite{wang2020network}, the prediction of CNN is used to feed a smaller MILP model. 

For these above works \cite{lei2019learning,lei2017deep,wang2019caching,wang2020network,lee2018deep}, the FNN or CNN input is extracted from the optimization model. The work in \cite{ling2020solving} trains a CNN via original graph instead of parameters matrix/image, which makes the process human recognizable and interpretable. Though the authors take traveling salesman problem not data caching as an example, the method can be viewed as a potential research direction. 

In \cite{doan2018content}, an ILP model is proposed to minimize the backhaul video-data type load by determining the portion of cached content in BSs. Considering the fact that the mobile users covered by a BS change frequently, therefore predicting user preference is unnecessary. Instead, the authors concentrate on the popular content in general. At the beginning, a 3D CNN is introduced to extract spatio-temporal features of videos. The popularity of new contents without historical information is determined via comparing similar video features. The authors of \cite{qin2019neural} also considers the spatio-temporal features among visiting contents in a mobile bus WiFI environment. By exploiting the previous 9 days collecting data, the content that the user may visit on the last day and corresponding visiting frequency can be forecast. The social property is taken into account in \cite{tsai2018mobile}. By observing users interests on tweets during 2016 U.S. election, a CNN based predicted model can foresee the content category that is most likely to be requested. Such kind of content would be cached in MBSs and FBSs.

The work of \cite{fedchenko2019feedforward} examines the role of DNN in caching from another aspect. The authors propose a FNN to predict content popularity as a regression problem. The results show that FNN outperforms RNN, though the later is believed to be effective to solve sequential predictions. Moreover, replacing the FNN by a linear estimator does not devalue the performance significantly. The author provides explanation that FNN would work better than linear predictor in the case of incomplete information, and RNN has more advantages to model the popularity prediction as a classification rather than a regression problem. 

\subsection{RNN}

Considering RNN is superior in dealing with sequential tasks, the work \cite{ale2019online} applies a bidirectional RNN for online content popularity prediction in mobile edge network. Simple RNN's output depends on previous and current storage, but the bidirectional can also take future information into account. The forecast model consists three blocks cascadingly: a CNN reads user requests and extract features; bidirectional LTSM learns association of requests over time step; FNN is added in the end to improve the prediction performance. Then content eviction is based on the popularity prediction. 

The authors in \cite{chen2017echo} utilize ESN to predict both content request distribution and end user mobility pattern. The user's preference is viewed as context which links with personal information combining gender, age, job, location, etc.. For the request prediction, the input of ESN is user's information vector and the output represent the probability distribution of content. For mobility prediction, the input includes historical and present user's location and the output is the expected position for next time duration. Eventually, the prediction influences the caching content decisions in BBUs and RRHs for the purpose of minimizing traffic load and delay in CRAN. The authors extend their work in \cite{chen2017caching} by introducing conceptor-based ESN which can split users' context into different patterns and learn them independently. Therefore a more accurate prediction is achieved. 

In \cite{fan2020pa}, a caching decision policy named PA-Cache is proposed to predict time-variant video popularity for cache eviction when the space is full. The temporal content popularity is exploited by attaching every hidden layer representation of RNN to an output regression. In order to improve the accuracy, hedge backpropagation is introduced during training process which decides when and how to adapt the depth of the DNN in an evolving manner. Similarly, the work in \cite{zhang2019toward} also considers caching replacement of video content. A deep LSTM network is utilized for popularity prediction consisting of stacking multiple LSTM layers and one softmax layer, where the input of the network is request sequence data (device, timestamp, location, title of video) without any prepossessing and the output is estimated content popularity. Another work concentrates on prediction and interactions between user mobility and content popularity can be found \cite{mou2019lstm}.

The work of \cite{narayanan2018deepcache} recognizes the popularity prediction as a seq2seq modeling problem and proposes LSTM Encoder-Decoder model. The input vector consists of past probabilities where each vectors are calculated during a predefined time window. In \cite{zhang2019cover}, the authors focus on caching content delivery with the aim of minimizing BSs to cover all requested users, i.e. set cover problem, via coded caching. Unlike \cite{narayanan2018deepcache}, an auto encoder is introduced in coded caching stage for file conversion to reduce transmission load. In addition, a RNN model is employed to select BSs for broadcasting. 

The paper \cite{jiang2020neural} shows the potential of RNN in solving where to cache problem. In \cite{jiang2020neural}, a task allocation model is formulated as a knapsack problem and the decision variables represent the task is processed locally in mobile devices (MDs) or remotely in edge servers (ESs). The authors design a multi-pointer network structure of 3 RNNs, where 2 encoders encode MDs and ESs respectively, 1 decoder demonstrates ES and MD pairing. Considering the similarity of where to cache optimization model and knapsack problem, the multi-pointer network can be transferred for caching location decision after according parameter modifications.

\subsection{Auto Encoder}
Generally, auto encoder is utilized to learn efficient representation or extract features of raw data in an unsupervised manner. The work in \cite{lei2019deep} considers the cache replacement in wireless sensor network (WSN) based on content popularity. Considering sparse auto encoder (SAE) can extract representative expression of input data, the authors employ a SAE followed by a classifier where the input contains collecting user content requests and the output represents the contents popularity level. The authors also think about the implementation in a distributed way by SDN/NFV technical, i.e. the input layer is deployed on sink node, while the rest layers are implemented on the main controller. A related work applying auto encoder in 5G network proactive caching can be found in \cite{lei2018proactive}. In \cite{rathore2019deepcachnet}, two auto encoders are utilized for extracting the features of users and content respectively. Then the extracted information is explored to estimate popularity at the core network. Similarly, the auto encoder in \cite{liu2017content} is for spatio-temporal popularity features extraction and auto encoders work collaboratively in \cite{lin2020video} to predict top K popular videos.

\subsection{DRL}

The work in \cite{gao2020reinforcement} focuses on the cooperative caching policy at FBSs with maximum distance separable coding in ultra dense networks. A value-based model is utilized to determine caching categories and the content quantity at FBSs during off peak duration. The authors in \cite{maniotis2020viewport} study the problem of caching 360\degree videos and virtual viewports in FBSs with unknown content popularity. The virtual viewport represents the most popular tiles of a 360\degree video over users' population. A DQN is introduced to decide which tiles of a video to be hosted and in which quality. Additionally, \cite{he2019qoe} employs DQN for content eviction decision offering a satisfactory quality of experience and \cite{tang2019energy} is for the purpose of minimizing energy consumption. In \cite{wu2019dynamic}, the authors also apply DQN to decide cache eviction in a single BS. Moreover, the critic is generated with stacking LSTM and FNN to evaluate Q value and an external memory is added for recording learned knowledge. For the purpose of improving the prediction accuracy, the Q value update is determined by the similarity of estimated value of critic and recording information in the external memory, instead of critic domination. The paper \cite{sadeghi2019deep} puts forth DQN a two-level network caching, where a parent node links with multiple leaf nodes to cache content instead of a single BS. In \cite{zhong2018deep}, a DRL framework with Wolpertinger architecture \cite{dulac2015deep} is presented for content caching at BSs. The Wolpertinger architecture is based on actor-critic model and performs efficiently in large discrete action space. \cite{zhong2018deep} employs two FNNs working as actor and critic respectively, where the former determines requested content is cached or not and the later estimates the reward. The whole framework consists two phases: in offline phase, these two FNNs are trained in supervised learning; in online phase, the critic and actor update via the interaction with environment. The authors extend their work to a multi agent actor-critic model for decentralized cooperative caching at multiple BSs \cite{zhong2020deep}. In \cite{zhu2018caching}, an actor-critic model is used for solving cache replacement problem, which balance the data freshness and communication cost. The aforementioned papers put attention on the network performance while ignore the influence of caching on information processing and resource consumption. Therefore, authors of \cite{wang2019deepchunk} design cache policy considering both network performance during content transmission and processing efficiency during data consumption. A DQN is employed to determine the number of chunks of the requested file to be updated. The paper \cite{rahman2020learning} investigates a joint cache replacement and power allocation optimization problem to minimize latency in a downlink F-RAN. A DQN is proposed for finding a suboptimal solution. Though \cite{thar2019deep} is regarded as solving what to cache problem like \cite{zhong2018deep,zhong2020deep,gao2020reinforcement,maniotis2020viewport,he2019qoe,tang2019energy,wu2019dynamic,zhu2018caching,sadeghi2019deep,wang2019deepchunk,rahman2020learning}, the reinforcement learning approach plays a different role. In \cite{thar2019deep}, a DNN is utilized for content popularity prediction and then a RL is used for DNN hyperparameters tuning instead of determining caching content. Therefore the action space consists of choosing model architectures (i.e. CNN, LSTM, etc.), number of layers and layer configurations.

In \cite{he2018trust}, the authors generate an optimization model with the aim of maximizing network operator's utility in mobile social networks under the framework of mobile edge computing, in network caching and D2D communications (3C). The trust value which if estimated through social relationships among users are also considered. Then a DQN model is utilized for solving optimization problem, including determine video provider and subscriber association, video transcoding offloading and the video cache allocation for video providers.. The DQN employs two CNNs for training process, where one generates target Q value and the other is for estimated Q value. Unlike the conventional DQN, the authors in \cite{he2018trust} introduces a dueling structure, i.e. the Q value is not computed in the final fully connected layer, but is decomposed into two components and use the summary as estimated Q value, which helps achieve a more robust result. The authors also consider utilizing dueling DQN model in different scenarios like cache-enabled opportunistic interference alignment \cite{he2017deep} and orchestrating 3C in vehicular network \cite{he2017integrated}. The work \cite{qiao2019deep} provides a DDPG model to cope with continuous valued control decision for 3C in vehicular edge networks, which is combined with the idea of DQN and actor-critic model. The DDPG structure can be divided into two parts as DQN, one is for estimated Q value and the other for target Q value. Each part consists of two DNNs, which play the role of actor and critic respectively. The critic updates its parameters like DQN while the actor learns policy via deterministic policy gradient approach. The proposed DRL is used for deciding content caching/replacement, vehicle organization and bandwidth resource assignment on different duration. 

The paper \cite{wei2018joint} provides an optimization model which takes what to cache and content delivery into consideration in the fog-enabled IoT network in order to minimize service latency. Since the wireless signals and user requests are stochastic, a actor-critic model is engaged where the actor makes decision for requesting contents while critic estimates the reward. Specially, the action space $S$ consists of decision variables and reward function is a variant of the objective function. A caching replacement strategy and dynamic multicast scheduling strategy are studied in \cite{zhang2019double}. In order to get a suboptimal result, an auto encoder is used to approximate the state. Further, a weighted double DQN scheme is utilized for avoiding overestimation of Q value. \cite{li2020joint} applies a RNN to predict content popularity by collecting historical requests and the output represents the popularity in the near future. Then the prediction is employed for cooperative caching and computation offloading among MEC servers, which is modelled as a ILP problem. For the purpose of solving it efficiently, a multi-agent DQN is applied where each user is viewed as an agent. The action space consists of task local computing and offloading decision as well as local caching and cooperative caching determination. The reward is measured by accumulated latency. The agent choose its own action based on current state without cooperation. The where to cache, what to cache and content delivery decision of D2D network are jointly modelled in \cite{li2019deep}. Two RNNs, ESN and LSTM, are considered to predict mobile users' location and requested content popularity. Then the prediction result is used for determining content categories and cache locations. The content delivery is formulated as the actor-critic based DRL framework. The state spaces include CSI, transmission distances and communication power between requested user and other available candidates. The function of DRL is determining the communication link among users with the aim of minimizing power consumption and content delay.

We notice that most papers prefer to use value-based model (critic) and value-policy-based (actor-critic) model in DRL framework, but rare paper considers only policy-based model to solve data caching problem. One proper reason is that the search space of caching problem is enormous so policy-based model is easier to fall into local optimal solution, resulting in poor performance. Though the value-based model is less stable, some variant structures are utilized like Double DQN in \cite{zhang2019double} to avoid value overestimation and dueling DQN in \cite{he2018trust,he2017deep,he2017integrated} to improve robust.
\section{Research Challenges and Future Directions}
\label{sec:challenges}

A serious of open issues on content caching and potential research directions are discussed in this section. We first extend the idea of content caching to virtual network function chain since caching can be viewed as a specific network function. Then we consider the caching for augmented reality applications. Moreover, we notice that the cache dimensioning has not been covered yet by DL methods. Finally, we debate the addition cost introduced by DL.

\subsection{Caching as a Virtual Network Function Chain}
The concept of Network Function Virtualization (NFV) has been firstly discussed and proposed within the realms of the European Telecommunication Standardization Institute (ETSI)\footnote{Network Functions Virtualisation, An Introduction, Benefits, Enablers, Challenges and Call for Action, ETSI, 2012 https://portal.etsi.org/NFV/NFV\_White\_Paper.pdf}. The rational is to facilitate the dynamic provisioning of network services through virtualization technologies to decouple the service creation process form the underlying hardware. The framework allows network services to be implemented by a specific chaining and ordering of a set of functions which can be implemented either on a more traditional dedicated hardware which in this this case are called Physical Network Functions (PNFs), or alternatively as Virtual Network Functions (VNFs) which is a software running on top of virtualized general-purpose hardware. The decoupling between the hardware and the software is one of the important considerations the other – equally important – is that a virtualized service lend itself naturally to a dynamic programmable service creation where VNF resources can be deployed as required. Hence, edge cloud and network resource usage can be adapted to the instantaneous user demand whilst avoiding a more static over-provisioned configurations.

Within that framework, the incoming network service requests include the specification of the service function chain that need to be created in the form of an ordered sequence of VNFs. For example different type of VNFs such as a firewall or a NAT mechanism need to be visited in a specific order. In such  constructed service chain each independent VNF requires specific underlying resources in terms for example of CPU cycles and/or memory. 

Under this framework, caching of popular content can be considered as a specialized VNF chain function since inevitably delivery of the cached popular content to users will require a set of other functions to be supported related to security, optimization of the content etc. However, the issue of data caching and VNF chaining have evolved rather independently in the literature and the issue on how to optimize data caching when seeing it as part of VNF chain is still an interesting open ended issue. 

\subsection{Caching for Mobile Augmented Reality (MAR) applications and Digital Twins (DTs)}

Mobile  augmented reality (MAR) applications can be considered as a way to augment the physical real-world environment
with artificial computer-based generated information and is an area that has received significant research attention recently. In order to successfully superimpose different digital object in the physical world MAR applications include several computationally and storage complex concepts such as image recognition, mobile camera calibration, and also the use of advanced 2D and 3D graphics rendering. These functionalities are highly computationally intensive and as such require support from an edge cloud, in addition the virtual objects to be embedded in the physical world are expected to be proactively cached closer to the end user so that latency is minimized. Ultra low latency in these type of applications is of paramount importance so that to provide a photorealistic embedding of virtual objects in the video view of the end user. However, since computational and augmented reality objects need to be readily available, the caching of those objects should be considered in conjunction with the computational capabilities of he edge cloud. In addition to the above when MAR is considered under the lenses of an NFV environment the application might inherently require access to some VNFs and therefore the above discussion on VNF chaining for MAR applications is also valid in this case. 

Recently the concept of Digital Twin (DT) \cite{DTIEEEMult}, \cite{DTIEEEAccess} has received significant research attention due to the plethora of applications ranging from industrial manufacturing and health to smart cities. In a nutshell, a DT can be defined as an accurate digital replica of a real world object across multiple granularity levels; and this real world object could be a machine, a robot or an industrial process or (sub) system. By reflecting the physical status of the system under consideration in a virtual space open up a plethora of optimization, prediction, fault tolerance and automation process that cannot be done using solely the physical object. At the core of DT applications is the requirement of stringent two–way real time communication between the digital replica and the physical object. This requirement inevitably require support from edge clouds to minimize latency and efficient storage and computational resources including caching. In that setting, the use of the aforementioned deep learning technologies will have a key role to play in order to provide high quality real time decision making to avoid misalignment between the digital replica of the physical object under consideration. Efficient machine-to-DT connectivity would require capabilities similar to the above mentioned augmented reality application but due to the continuous real-time control-loop operation DTs will require a complete new set of network optimization capabilities and in that frontier efficient caching and data-driven techniques will have a central role to play. Hence, as the research regarding the inter-play between low latency communications and DTs is still in embryonic stage there is significant scope in the investigation of suitable data driven deep learning techniques to be utilized for distributed allocation of caching and computing resources.

\subsection{Deep Learning for Cache Dimensioning}

As introduced in Section \ref{sec:caching}, cache dimensioning explores the appropriate cache size allocation for content host such as CRs and BSs. Disappointingly, there is rare paper applies DL on cache dimensioning decisions. One proper reason is lack of training data set in contrast to content popular prediction, where we have historical user request log to train a DNN. In addition, the caching size allocation affects the network performance and economic investment. Recently, network slicing is identified as an important tool to enable 5G to provide multi-services with diverse characteristics. The slice is established on physical infrastructure including network storage. Therefore it is a very interesting topic to consider the allocation of the memory space to support content caching and other storage services, which guarantees QoE and satisfies task requirements. Furthermore, for the case lack of training data set, DRL can be viewed as a promising technology to configure slicing settings as well as cache dimensioning. For the action space designing, it can be either discrete by setting storage levels, or continuous which is allocate the memory space directly. However, there are requirements to design caching-enabled network slicing model especially for dynamic allocation as well as associated DRL framework including state space, detailed action space, reward function and agent structure.

\subsection{The Cost of Deep Learning}
Though the application of DL brings performance efficiency for caching policy, additional cost introduced by DL is unneglected, since training and deploying DL model require not only network resources but also time duration. Naturally there is a trade-off between the cost which DL-assisted caching policy saved and the consumption which supports DL itself running, which indicates the trading with DL results in either profit, loss, or break even. Therefore, where and when to apply DL should be carefully investigated. In addition, for the purpose of reducing resource consumption and accelerating training process, some knowledge transfer methods like transfer learning \cite{weiss2016survey} can be utilized, which can transform the knowledge already learnt from the source domain to a relevant target domain.

\section{Conclusions}
\label{sec:conclusions}

This article presents a comprehensive study for the application of deep learning methods in the area of content caching. Particularly, the data caching is divided into two classifications according to the caching location of edge network. Each category contains where to cache, what to cache, cache dimensioning and content delivery. Then we introduce typical DNN methods which are categorised via training process into supervised learning, unsupervised learning and RL. Further, this paper critically compares and analyzes state-of-the-art papers on parameters, such as DL methods employed, the caching problems solved and the objective of applying DL. The challenges and research directions of DL on caching is also examined on the topic of extending caching to VNF chains, the application of caching for MAR as well as DTs, DL for cache size allocation and the additional cost of employing DL. Undoubtedly, DL is playing a significant role in 5G and beyond. We hope this paper will increase discussions and interests on DL for caching policy design and relevant applications, which will advance future network communications. 

\vspace{6pt} 

\bibliographystyle{unsrt}
\bibliography{reference}

\end{document}